\documentstyle[12pt,epsfig]{article}
\begin{document}
\begin{center}
\vspace{1.5in}
{\LARGE
Triton clustering in neutron rich nuclei and Ikeda - like diagrams }
\end{center}
\vspace{.4in}
\begin{center}
{\bf Afsar Abbas}\\
\vspace{.1in}
Institute of Physics\\ 
Bhubaneshwar-751005, India\\
\vspace{,1in}
email: afsar@iopb.res.in
\end{center}
\vspace{1.5in}
\begin{center}
{\bf Abstract}
\end{center}
\vspace{.3in}

It is shown here that new experiments confirm author's recent prediction 
of a strong tendency for triton clustering in light neutron rich nuclei.
As such the neutron halo phenomena is naturally explained here.
Prediction of exotic molecular states rich in tritonic clusters is made
here through new Ikeda-like diagram configurations for the nuclei 
${^{3Z}_Z} A _{2Z}$ . Hence this model is well confirmed experimentally 
and makes several new predictions which can be tested experimentally.

\newpage

Clustering , in particular that of $\alpha$, has been a persistent aspect 
of nucler structure and reaction studies [1]. So much so that some have 
speculated of the existence of $\alpha$ - matter in contrast to nuclear 
matter. Hence it is commonly believed that multi-$\alpha$ structures of 
various geometry should be present in nuclei [2]. In fact even linear 
$\alpha$ chains are expected to occur too [3,4,5]. Even when the neutron 
number increases drastically ( as it happens in the case of neutron rich 
light nuclei ) the $\alpha$ structure is believed to persist [1,2]. 
This is the standard picture of neutron rich nuclei accepted today. 

However it may very well be that as neutron number increaes away from the 
line of nuclear stability, some new structures may start manifesting 
themselves. Indeed a new model arising from Quantum Chromodyanamics (QCD)
and quark model considerations shows that this actually should be 
happening [6]. Therein the author has recently predicted that 
triton clustering rather than $\alpha$ clustering should be the dominant 
aspect of the structure of neutron rich nuclei [6]. In this paper
here, experimental evidences shall be presented to support this view of 
triton clustering in neutron rich nuclei. In addition Ikeda - like 
diagrams with triton rich linear structures ( which can be tested 
experimenatlly ) shall also be presented. One interesting case where one 
such prediction has already been confirmed shall be discussed.

Ideas based on QCD and quark model allow us to see a common thread 
in such diverse nuclear phenomenon as [6]: 
1. The hole at the centre of the density distribution in ${^3}H, {^3}He$ 
and ${^4}He$,
2. The halo phenomena is neutron rich nuclei,
3. The formation and persistence of clusters in nuclei and
4. The nuclear molecule effects.
It turns out that all these require an understanding of two or more 
nucleons strongly overlapping over a small local region of size
$\le 1 fm$. This necessarily requires considerations of multi-quark 
configurations like 6-, 9- and 12-quarks. It was shown by the author
[6,7,8] that the nucleons do not like to go into these configurations and 
thereby the above mentioned nuclear effects are given a consistent 
explanation. The reader may refer to [6,7,8] for further details.
  
For our purpose here, note the effect of triton clustering for neutron 
rich nuclei predicted in this model [6]. It was shown that as more and 
more neutrons are added to light nuclei there is a marked effect that an 
$\alpha$ plus two neutrons prefer to go to a configuration of two tritons.
There is a continuous competition between an $\alpha$ cluster plus two 
neutrons remaining as such versus these changing into two clusters of 
tritons. For a single such configuration in 
${^6}He$ this breakup does not happen and to a 
very good approximatioon ${^6}He$ remains an $\alpha$ plus two neutrons 
for the ground state and low energy excitations. This has important 
consequences like forming a halo [9,10] and thus providing us with 
a better insight into the structure of other A=6 nuclei like
${^6}Li$ and ${^6}Be$ [11]. The effect of triton clustering also 
helps us to explain the phenomenoa of neutron halos [6].

In our model for example, ${^{12}}Be$ rather being treated as 
consisting of 2$\alpha$ + 4n configuration ( as done by others [1,2] ) 
is treated as made up of 4t ( note that 't' stands for 
triton ) [6]. As such it forms a compact core and when two more 
neutrons are added to it they are consigned to stay outside the 
surface of the core and thereby forming a halo for ${^{14}}Be$.
It is clustering of either only tritons or with tritons and $\alpha$
that one gets a compact core, like ${^{10}}Be$ froms a core of
$\alpha$ + 2t to give ${^{11}}Be$ a one neutron halo structure [6].

${^{18}}C$ is made up of 6t and then ${^{19}}C$ forms a one neutron halo
around this core. ${^{17}}B$ is two neutron halo nucleus with a core 
of ${^{15}}B$ made up of 5t. This structure is indeed so stable that it 
can accept four more neutrons attached to its surface to form ${^{19}}B$.

There have been claims that inspite of higher values for 
one neutron and  two neutron
separation energies for ${^{15}}C$ and ${^{16}}C$ respectively, these 
are actually one and two neutron halo nuclei [12]. In our model ${^{14}}C$ 
forms a compact core of the structure 2$\alpha$ + 2t. Our model also 
makes a unique prediction that this nucleus should have marked 
depression at the centre in its matter distribution and also therefore 
have a pronounced excess of matter concentration on the surface [6]. 
(It is hoped that with the currently
proposed lepton-RIB colliders it should be possible to obtain charge 
density distribution of these radioacutive nuclei in the near future). 
Thereby the extra neutrona in ${^{15}}C$ and ${^{16}}C$ would not be
able to penetrate through the surface as that would entail 6-,9- or 
12-quark configurations locally and as this is suppressed due to QCD and 
quark model considerations, these extra neutrons form a halo around this 
core of ${^{14}}C$.

As per this model ${^{21}}N$ would consist of 7t and form a good core.
Then ${^{22}}N$ and ${^{23}}N$ would appear as one-neutron and two-neutron 
halo nuclei respectively - exacly as seen in the experiments [9].
For the Oxygen case ${^{24}}O$ is made up of 8t and in our model this 
should be extra stable. Interestingly this is what has been found 
experimentally [9]. 
Note that in our model ${^{22}}O$ has structure $\alpha$ + 6t.
The number of clusters here are the same as in the very stable 7t 
structure of ${^{21}}N$ as indicated above. Hence this nucleus could 
also be very stable. Thus as one or two neutrons are added to it, it may 
show them as halo nuclei. Thus both ${^{23}}O$ and ${^{24}}O$ would appear 
as halo nuclei [9]. This interesting situation calls for further study. 
However, first it has to be clearly understood whether it is a real halo 
effect we are seeing or a very thick neutron skin.

It was suggested by Ikeda [13] that cluster like configurations would 
appear in light A=4n self conjugate nuclei near the threshold energy for 
breakup into proper sub-nuclei. The corresponding Ikeda diagrams gave 
visual pictures of relevant structures [1,2]. Note that this necessarily 
involved clusters of $\alpha$ particles.

In our model here as triton clustering appears to be such a dominant 
effect for neutron rich nuclei [6], then quite clearly we should expect that 
for nuclei ${^{3Z}_Z} A _{2Z}$ similar triton rich cluster structures 
would exist. Therfore configurations with neutron rich nuclei would 
appear near the threshold energy for decay into corresponding relevant 
sub-nuclei. These Ikeda - like diagrams are here given in Figure 1.  
So as not to clutter the figure not all the relevant diagrams are shown 
here. For example for ${^{12}}Be$ the configuration ${^6}He$ + 2t ( 22.42 
MeV) is not shown. Also not listed are situations with one or more 
$\alpha$ in it eg. $\alpha$ + 2t + 2n for ${^{12}}Be$.

Recently exotic molecular states in ${^{12}}Be$ have been found [14].
This was in the study of breakup of ${^{12}}Be$ into  ${^6}He$ + ${^6}He$ 
and ${^4}He$ + ${^8}He$. These can be interprested as coming from
$\alpha$ - 4n - $\alpha$ configurations [1,14]. This is the standard way. 
However these same molecular resonances can also 
be understood in our model in terms of 
Ikeda-like diagrams given here ( I thank Isao Tanihata for bringing this 
experiment to my attention [15] ). Thus we can treat this experiment to be 
confirming our prediction here also. However, unique confirmation 
would come when molecular configurations rich in tritons, like 
for example the existence of molecular configurations ${^9}Li$ +t for 
${^{12}}Be$ breakup would be demonstrated. These experiments should be 
possible with present day techniques with RIB in different laboratories.

In summary, further experiments continue to confrm author's recent model 
prediction of strong triton clustering tendencies in neutron rich 
nuclei. This helps to explain the neutron halo phenomena also. This 
allows us to make predictions for triton rich molecular configurations 
arising from Ikeda - like diagranms for the neutron rich nuclei. 

\newpage

\vspace{.5in}
Acknowledgement:

The author would like to thank Dr. P. Arumugam for help with the figure.


\vspace{.2in}

{\bf References} 

\vspace{.2in}

1. W von Oertzen, Prog. Theor. Phys. Suppl. {\bf 146} (2002) 171

\vspace{.1in}

2. W von Oertzen, Z. Phys. {\bf A 357} (1997) 355

\vspace{.1in}

3. N. Orr, Eur. Phys. J., {\bf A 15} (2002) 109
 
\vspace{.1in}

4. K. Riisager, Rev. Mod. Phys. {\bf 66} (1994) 1105

\vspace{.1in}

5. L. Zamick and D. C. Zheng, Z. Phys. {\bf A 349} (1994) 255 

\vspace{.1in}

6. A. Abbas, Mod. Phys. Lett. {\bf A 16} (2001) 755.

\vspace{.1in}

7. A. Abbas, Phys. Lett. {\bf B 167} (1986) 150. 

\vspace{.1in}

8. A. Abbas, Prog. Part. Nucl. Phys. {\bf 20} (1988) 181.

\vspace{.1in}

9. I. Tanihata. Nucl. Phys. {\bf A 685} (2001) 80c

\vspace{.1in}

10. P. G. Hansen, Nucl. Phys. {\bf A 553} (1993) 89c

\vspace{.1in}

11. A. Abbas, " Structure of A=6 nuclei: ${^6}He$, ${^6}Li$ and ${^6}Be$,

IOP Preprint: IP/BBSR/2003-18 (June '03): physics/0306186

\vspace{.1in}

12. T. Zheng et.al., Nucl. Phys. {\bf A 709} (2002) 103

\vspace{.1in}

13. K. Ikeda, N. Takigawa and H. Horiuchi, Suppl. Prog. Phys. (Japan)
Extra (1969) 464

\vspace{.1in}

14. M. Freer et. al., Phys. Rev. Lett. {\bf 82} (1999) 1383

\vspace{.1in}

15. I. Tanihata - private communication

\newpage

\begin{figure}
\caption{Ikeda - like diagrams for neutron rich nuclei ${^{3Z}_Z} A
_{2Z}$.}
\epsfclipon
\epsfxsize=0.99\textwidth
\epsfbox{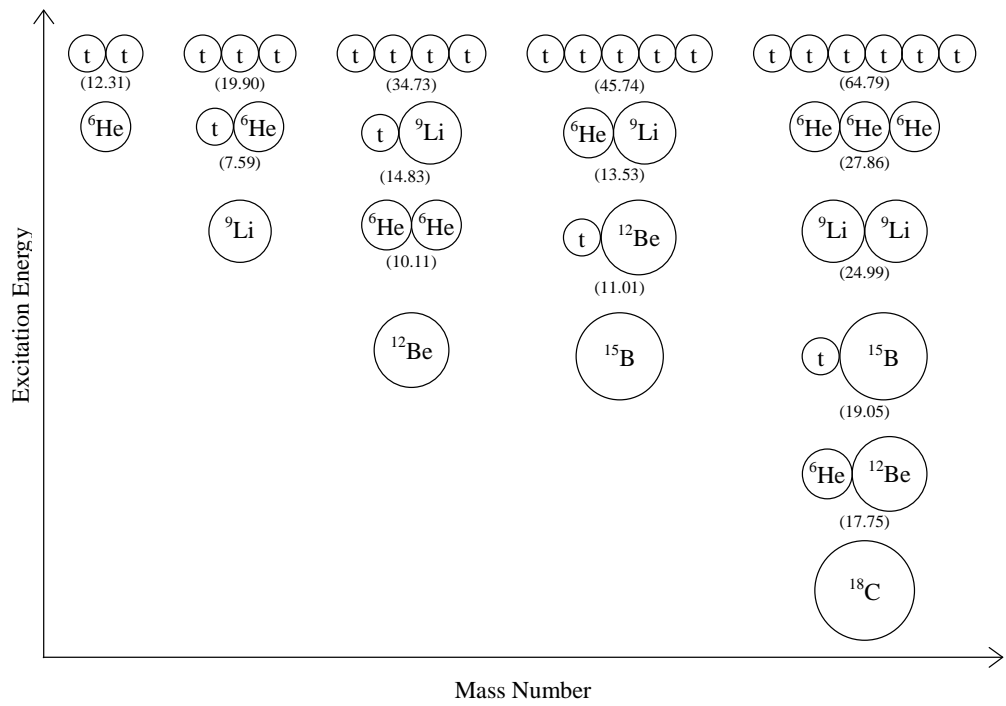}
\end{figure}

\end{document}